# High-Resolution Imaging of Plant Delayed Luminescence

Yan-Xia Liu[1], Hai-Yu Fan[1], Yu-Hao Wang[1,4], Yan-Liang Wang[2], Sheng-Wen Li[1], Shi-Jian Li[3, *], Xu-Ri Yao[1, *], Qing Zhao[1, *]

*1Center for Quantum Technology Research, School of Physics, Beijing Institute of Technology, Beijing 100081, China*

*2The Germplasm Bank of Wild Species & Yunnan Key Laboratory for Fungal Diversity and Green Development, Kunming Institute of Botany, Chinese Academy of Sciences, Kunming, Yunnan 650201, China*

*3School of Integrated Circuits and Electronics, Beijing Institute of Technology, 100081 Beijing, China*

*4Kunming Institute of Physics, 650221 Kunming, China*

*\*Author for correspondence: shijianli@bit.edu.cn (Shi-Jian Li), yaoxuri@bit.edu.cn (Xu-Ri Yao), qzhaoyuping@bit.edu.cn (Qing Zhao)*

Abstract

Delayed luminescence (DL) is a quantized signal that is characteristic of photoexcited molecules entering a relaxed state. Studying DL provides critical insight into photophysical mechanisms through the analysis of specific spatiotemporal dynamics. In this study, we developed a high-sensitivity DL imaging system using a quantitative scientific complementary metal-oxide-semiconductor (qCMOS) camera and a single-photon counting resolution. By optimizing the optical architecture and signal processing algorithms together, we achieved full-field spatiotemporal DL imaging at megapixel resolution (i.e., 2304 × 4096 pixels). Key findings include the following: (1) we observed spatial heterogeneity in DL intensity across the leaves of *Arabidopsis thaliana*, with stronger signals detected in veins and at sites of mechanical injury; (2) species-specific DL responses occur in response to oxidative stress, with *Hydrocotyle vulgaris* and *Ginkgo biloba* showing enhanced central DL activity; (3) excitation using white light induced maximum DL intensity, while red and blue light differentially modulated decay kinetics. Finally, we develop a two-level quantum model that links DL dynamics to the populations of excited-state electrons, thereby developing a theoretical framework for future photophysical research. Collectively, this work establishes a theoretical and technological framework for advancing plant phenotyping under stress conditions and optimizing light environments.

## 1. Introduction

Ultra-weak photon emission (UPE), a form of radiation that is intrinsically produced by living systems, exhibits photon fluxes ranging from several to hundreds of photons/(cm²·s), which encode critical information regarding biological redox states[1-2]. Delayed luminescence (DL) is a core component of UPE and originates from the non-radiative transitions of photoexcited molecules. Moreover, its kinetic characteristics are intrinsically linked to metabolic networks related to reactive oxygen species (ROS)[3-5]. The spatiotemporal distribution of DL is a rich source of information regarding quantum relaxation. This is because spatial gradients in DL intensity directly reflect the spatial occupancy profiles of excited-state electrons, while decay kinetics are highly correlated to energy-level transition rates.

ROS function as crucial signaling molecules that mediate rapid plant responses to biotic and abiotic stimuli, playing a pivotal role in growth, development, and environmental adaptation[6]. Under physiological homeostasis, ROS production and scavenging are maintained in a dynamic equilibrium via antioxidant systems, resulting in a stable baseline DL output. However, mechanical injury or oxidative stress can induce spatiotemporal ROS accumulation and amplify DL intensity via cascade effects. DL may be a highly sensitive biomarker for the plant stress response[7-12].

Fundamental biochemical studies confirm that DL originates from de-excitation of photoexcited species generated through radical cascades, primarily singlet oxygen ($^1O_2$) and excited carbonyls, rather than direct chlorophyll fluorescence[13-15]. While chlorophyll serves as the primary energy harvester in photosynthesis, its role in DL is indirect: functional chloroplasts facilitate reverse electron transfer between primary acceptors and oxidized chlorophyll, populating excited singlet states that emit photons during relaxation[13,16]. Significantly, this process requires intact photosynthetic machinery: inactivated chlorophyll or isolated pigment solutions exhibit no detectable DL, which confirms that DL dynamics encode real-time photobiochemical activity rather than passive pigment distribution[14,17]. While chlorophyll concentration modulates DL dynamics, the primary objective of this study was to establish a high-resolution imaging methodology for spatiotemporal mapping of ROS-driven DL, distinct from dissecting chlorophyll-specific mechanisms, which remain a focus for future investigations.

Current DL detection technologies face three major challenges:

(1) Single-point detection limitations: This is due to the fact that the conventional photomultiplier (PMT), while offering single-photon sensitivity, lacks sufficient spatial resolution to resolve DL heterogeneity in biological tissues[17-19].

(2) Temporal resolution mismatch: Many approaches use an electron-multiplying charge-coupled device (EMCCD). Such device, despite enabling spatial imaging, exhibit poor short-term response capabilities, and this issue conflicts with DL's rapid decay kinetics (i.e., ranging from milliseconds to seconds)[20-21].

(3) Noise and quantum efficiency trade-offs: Intensified charge-coupled device (ICCD) and scientific complementary metal-oxide-semiconductor (sCMOS) cameras are constrained by low quantum efficiency (QE$\leqslant$30%) and severe noise interference ($\geqslant$1.2 e$^-$/pixel), respectively. These effects can obscure the spatial features of weak DL signals[22-24].

These technical bottlenecks severely hinder current-generation spatiotemporal heterogeneity analyses and subsequent quantitative modeling of DL. Traditional DL detection, which is limited to point-based modes and may have low spatial resolution (typically 256 × 256 pixels), can struggle to resolve the spatiotemporal dynamics of DL in biological tissues.

Recent breakthroughs in quantitative complementary metal-oxide-semiconductor (qCMOS) sensors address critical limitations in plant biophoton imaging. With sub-electron read noise (0.27 e⁻RMS) and photon-number-resolving (PNR) capability, qCMOS discriminates single-photon events in single pixels – overcoming EMCCD's binary restriction and multiplicative noise[25-26]. Its 4.6-μm pixels enable micron-scale resolution at 22 fps, maintaining superiority even after 8×8 binning (effective pixel size 36.8 μm) [27-28]. These advances permit simultaneous quantification of photon statistics and spatial heterogeneity, as demonstrated in quantum spatial correlation measurements[28] and high-flux (>1 photon/pixel) biosensing[25]. qCMOS excels at capturing rapid time-varying signals like the early-phase DL decay kinetics (0-30s). EMCCD remains well-suited for integrating very weak, steady-state signals over longer durations. This fundamental difference in temporal response capability makes qCMOS the superior choice for studying the dynamic processes of plant DL addressed in this work. Here, we deploy qCMOS technology for the first time in plant delayed luminescence studies, leveraging its PNR capacity and spatiotemporal precision to decode ROS dynamics.

By leveraging the high signal-to-noise ratio (SNR) photon imaging capability of the qCMOS under short exposure times, we conducted imaging experiments using leaves sampled from *Arabidopsis thaliana*, *Hydrocotyle vulgaris*, and *Ginkgo biloba*. This system achieved DL photon images with resolutions up to 2304 × 2304 pixels (30 s temporal resolution, photons/pixel/30s). Next, we used diverse observational and stimulation experiments to identify species-specific DL spatiotemporal distribution patterns, spatial specificity during the stress response, and dynamic trends that emerge under stress. Finally, we propose a two-level quantum model to unify these observations. Overall, this study establishes a novel theoretical framework and technological platform for evaluating plant stress resistance and optimizing light environments.

## 2. Materials and Methods

## 2.1 Plant Materials and Growth Conditions

Plant materials

*Hydrocotyle vulgaris*: This species has thin leaves (thickness 0.2 ± 0.05 mm), is highly sensitive to the presence of ROS, and is therefore suitable for the dynamic observation of mechanical injury and oxidative stress.

*Arabidopsis thaliana*: This species has a well-annotated genome and is therefore ideal for analyses of the molecular mechanisms involved.

*Ginkgo biloba*: This species has thick leaves (i.e., 0.8 ± 0.1 mm), strong antioxidative capacity, suitable for light wavelength and long-term stress studies.

Cultivation conditions

Next, we cultivated each of the test species as described below.

*Hydrocotyle vulgaris*: Plants were grown in an indoor growth chamber (light intensity 50 µmol·m$^{-2}$·s$^{-1}$, 12-hour light/dark photoperiod) at 25 ± 2°C and were watered three times per day.

*Arabidopsis thaliana*: Surface-sterilized seeds were first sown on 1% (w/v) agar medium, stratified at 4°C for three days, germinated in a growth chamber for five days, then transplanted into an autoclaved artificial substrate containing peat, vermiculite, and perlite (in a ratio of 1:1:1, v: v: v). Seedlings were then left to grow for three weeks. Growth conditions: 16-hour light/8-hour dark photoperiod (photon flux density 100 µmol·m$^{-2}$·s$^{-1}$), 25°C, 60% relative humidity.

*Ginkgo biloba*: Mature leaves were collected from healthy trees growing at the Beijing Institute of Technology campus. At collection, leaves were screened for intact veins and absence of pests/diseases, transported in humidity-controlled boxes to the lab, then dark-adapted for two hours before experimentation to eliminate the effects of prior light exposure.

## 2.2 DL Imaging System

For two-dimensional DL imaging (**Figure 1**), a qCMOS camera (Hamamatsu ORCA-Quest) was served as the core detector. This camera was equipped with a 25 mm Computar M2514-MP2 objective lens and operated at −40°C to suppress thermal noise. During experiments, the camera was housed in a laboratory dark room (20°C) and controlled externally via a computer. Imaging was performed at a resolution of 2304 × 4096 pixels in the photon number resolving (PNR) mode for single-pixel photon counting. To overcome the inherent challenge of detecting ultra-weak DL signals, we implemented pixel binning – a standard approach in low-light imaging to optimize the SNR while preserving spatial information. This technique physically combines charge signals from adjacent pixels during readout, defined by a binning factor (N = 1, 2, 4, or 8). The process follows the fundamental principle: SNR scales with $\sqrt{N}$ under read-noise-limited conditions. For high-signal regions, native 2304×2304 resolution (4.6 µm pixels) resolved subcellular features. In lower-signal regimes: whole-leaf heterogeneity mapping used N=2 binning (effective pixel size 9.2 µm), temporal kinetics analysis employed N=8 binning (effective pixel size 36.8 µm) to achieve 64× SNR gain . Pixel binning (N=8) combined signals from adjacent pixels during readout, following the SNR-scaling principle . This approach optimized the detection of ultra-weak DL signals while maintaining superior resolution compared to EMCCD systems.

After imaging, all raw data underwent the following standardized processing steps:

(1) Background correction: Pre-acquired dark fields (i.e., 3 × 30s exposures) were used to establish a noise baseline that was then subtracted from other raw images.

(2) Spatial denoising: A six-frame sequence (i.e., 0–180 s decay, 30 s per exposure) underwent 3 × 3 pixels median filtering, while retaining the median intensity values within the central pixel neighborhood.

(3) Signal normalization: Leaf region masks extracted valid pixels, with the photon flux calculated in photons/pixel/30s.

(4) Spatiotemporal analysis: Time-aligned frames were used to generate pixel-wise decay curves and were then used to quantify DL intensity gradients and decay rates. Time-aligned frames were processed to extract DL kinetic profiles through the following steps:

a. Leaf region photon flux calculation:

For each exposure frame (30-s interval), the total photon count $P(t)$ within the leaf mask was summed and normalized by the valid pixel area $A_{pixel}$ and exposure time:

$I(t) = \frac{P(t)}{A_{pixel} \times N^2}$ [photons/pixel/30s], where $t$= 30, 60, 90, 120, 150, and 180 s, respectively.

b. Temporal decay curve generation:

DL intensity values at the six time points {$I(0)$, $I(30)$, ... , $I(180)$} were plotted as a function of time.

c. Key parameter quantification:

Initial intensity: $I(0)=I(t=30)$, Residual intensity: $I_{res}=I(180)$, Normalized residual: $Res_\% = \frac{I(180)}{I(0)} \times 100$.

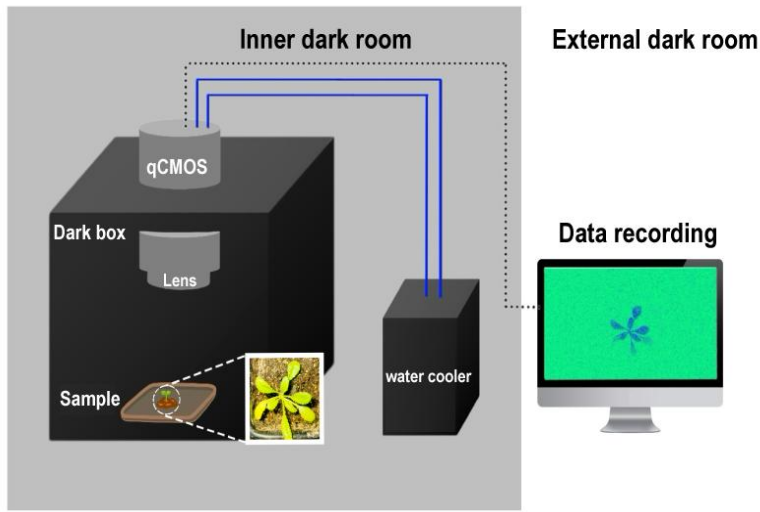

Figure 1. Diagram of the qCMOS imaging system.

## 2.3 Stress Treatments

Natural state imaging: DL imaging of intact *Arabidopsis thaliana* plants and leaves, *Hydrocotyle vulgaris* leaves, and *Ginkgo biloba* leaves was conducted under physiological equilibrium.

Stress treatments: All stress experiments used detached leaves. Pre-treatment: Leaves were dark-adapted for two hours in humidity-controlled boxes to eliminate the effects of prior light stimulation. During treatment, detached leaves were flattened in Petri dishes lined with moist filter paper and secured along petioles and margins using sterile double-sided tape.

Mechanical stimulation: A sterile blade was used to create a 2-cm incision at the midvein. Post-injury, leaves were immediately irradiated with white LED light (50 μmol·m$^{-2}$·s$^{-1}$) for five minutes to activate the photosynthetic system. DL signal acquisition began immediately after treatment, with 30-s exposures recorded continuously over a period of 180 s. The 5-min post-stimulation time point (representing peak DL response in initial trials) was selected for triplicate biological validation across all species.

Oxidative stress: Groups of plants were uniformly sprayed with 3% $H_2O_2$ solution (treatment) or deionized water (negative control). After incubation for two minutes to allow for solution penetration, leaves were then irradiated with white LED light (50 μmol·m$^{-2}$·s$^{-1}$) for five minutes, after which DL signal acquisition (30 s exposure) was performed. The 30-min post-$H_2O_2$ treatment time point (reflecting significant DL divergence from controls) was validated via three biological replicates.

Light wavelength excitation: Detached leaves were irradiated for five minutes with white (full spectrum), red (660 nm), or blue (460 nm) LED light (50 μmol·m$^{-2}$·s$^{-1}$). DL acquisition (using 30 s exposure times) commenced immediately after light cessation. The treatment order was randomized to avoid photoadaptation effects. Three independent leaves were tested per wavelength condition.

Experimental continuity protocol: For sequential stress-time experiments, each time point was assessed in an independent trial. Prior to each new stimulation, leaves underwent a 2-h dark adaptation followed by 5-min white-light re-irradiation to ensure complete decay of residual DL signals and re-establishment of baseline physiological conditions.

## 3. Results

## 3.1 High-Resolution Imaging Reveals DL Spatial Heterogeneity

Using a qCMOS-based high-resolution imaging system, we first conducted DL imaging experiments on unstressed plant leaves. **Figure 2** illustrates high-resolution DL images (N = 2, 1152 × 1152 pixels) of different plant leaves captured over the 0–180 s timeframe.

The temporal evolution of DL signals exhibited a biphasic pattern characterized by an initial rapid decline followed by asymptotic stabilization. During the initial excitation phase (0–30 s), DL signal intensities of all tested plant leaves peaked, with leaf morphology visible in DL images. Moreover, we observed species-specific kinetic characteristics in the subsequent decay phase. For example, *Arabidopsis thaliana* leaves exhibited rapid DL attenuation within 30–60 s, approaching baseline levels by 120 s. In contrast, *Ginkgo biloba* leaves displayed slower decay rates, retaining detectable intensities after 90–120 s, and *Hydrocotyle vulgaris* leaves showed intermediate decay kinetics, showing higher signal intensities during the early decay phase (30–90 s) and gradual convergence to intrinsic ultraweak luminescence levels after 150–180 s.

Distinct spatial heterogeneity in DL signal distribution was observed across the species examined. In *Arabidopsis thaliana* leaves, DL intensity was found to be significantly higher in vein regions relative to mesophyll tissue. In contrast, *Ginkgo biloba* leaves exhibited a centrally symmetric DL distribution, with intensity maxima localized to the leaf midzone. Different still were *Hydrocotyle vulgaris* leaves, which displayed a homogeneous, diffuse DL pattern. During the initial phase (0–30 s), DL images preserved full leaf contours (e.g., lamina, petiole, etc.), with luminescent regions precisely aligned to anatomical structures. During the decay phase (30–180 s), the basic leaf morphology remained discernible, with luminescence intensities declining synchronously across all regions.

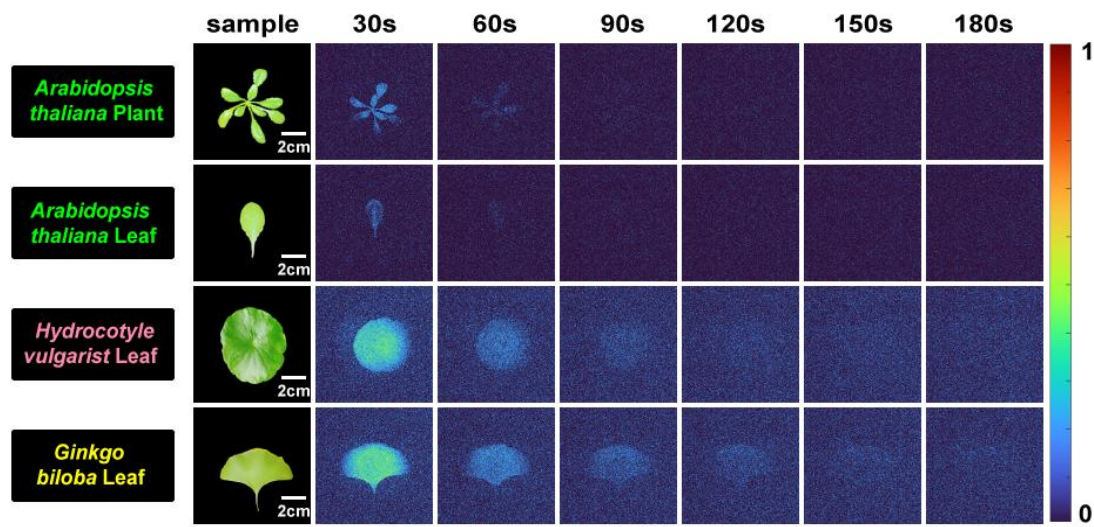

Figure 2. Spatiotemporal characteristics of delayed luminescence (DL) in three plant species under natural conditions. Pseudo-color heatmaps show DL intensity distributions in Arabidopsis thaliana plants/leaves, Hydrocotyle vulgaris leaves, and Ginkgo biloba leaves (image resolution: 1152 × 1152 pixels, achieved by binning N=2 pixels).

## 3.2 Wavelength-Dependent DL Modulation

Next, leveraging the broad spectral response (400–1000 nm) of the qCMOS imaging system, we systematically analyzed the regulatory effects of white, red (660 nm), and blue (460 nm) light excitation on DL signals in the leaves of *Arabidopsis thaliana*, *Ginkgo biloba*, and *Hydrocotyle vulgaris*, as shown in **Figure 3**.

Pseudo-color heatmaps (**Figure 3a–c**) depict the spatial distribution of DL intensity under different wavelengths. For example, *Arabidopsis thaliana* leaves showed the strongest DL under white light ($I_0$=0.32 photons/pixel/30s), while red ($I_0$=0.18 photons/pixel/30s) and blue light ($I_0$=0.17 photons/pixel/30s) induced weaker responses. The red/blue intensity ratio was approximately 1:1.

*Hydrocotyle vulgaris* leaves exhibited high-intensity signals under white light ($I_0$=0.50 photons/pixel/30s) and red light ($I_0$=0.44 photons/pixel/30s), both significantly stronger than blue light ($I_0$=0.27 photons/pixel/30s), yielding a red/blue ratio of 1.63:1. *Ginkgo biloba* showed maximal DL under white light ($I_0$=0.55 photons/pixel/30s) with a fan-shaped spatial distribution (stronger near midveins). Under red ($I_0$=0.32 photons/pixel/30s) and blue light ($I_0$=0.30 photons/pixel/30s), intensities were lower and the fan pattern less distinct.

Distinct species-specific response modes were also observed under different light qualities(**Figure 3d–e**). Overall, *Hydrocotyle vulgaris* leaves exhibited the strongest DL persistence under red light. The fan-shaped DL distribution of *Ginkgo biloba* leaves exposed to white light closely matched the leaf anatomy of this species, whereas this structural correlation diminished under monochromatic light. Finally, *Arabidopsis thaliana* leaves maintained significantly higher residual intensities under red/blue light ($Res_\%$>77%) than under white light ($Res_\%$=46.8%), showing lower signal variability.

Temporal dynamic analysis revealed species-specific DL decay patterns under different light qualities (**Figure 3f–h**). For *Arabidopsis thaliana* leaves, red light excitation generated the lowest initial DL intensity ($I_0$=0.17±0.0108 photons/pixel/30s) but the highest residual percentage ($Res_\%$=78.6). In contrast, blue light showed similar $I_0$ (0.17 ± 0.0033 photons/pixel/30s) with slightly lower $Res_\%$=77.9, while white light yielded the highest $I_0$ (0.29±0.0509 photons/pixel/30s) but rapid decay ($Res_\%$=46.8). For *Hydrocotyle vulgaris* leaves, blue light excitation generated the slowest decay ($I_{res}$=0.14±0.0009 photons/pixel/30s, $Res_\%$=54.4) despite low initial DL intensity ($I_0$=0.25±0.0177 photons/pixel/30s). In contrast, red light excitation yielded moderate $I_0$ (0.43 ± 0.0527 photons/pixel/30s) with $Res_\%$=32.6, while white light showed the highest $I_0$ (0.43 ± 0.0658 photons/pixel/30s) and fastest decay ($Res_\%$=32.0). For *Ginkgo biloba* leaves, white light excitation generated the highest initial DL intensity ($I_0$=0.50±0.0280 photons/pixel/30s), but also the fastest decay rate ($I_{res}$=0.14±0.0005 photons/pixel/30s, $Res_\%$=26.9). In contrast, red light excitation yielded lower $I_0$ (0.34±0.0196 photons/pixel/30s) but higher $Res_\%$=41.6, while blue light showed $I_0$=0.28±0.0183 photons/pixel/30s with $Res_\%$=48.5.

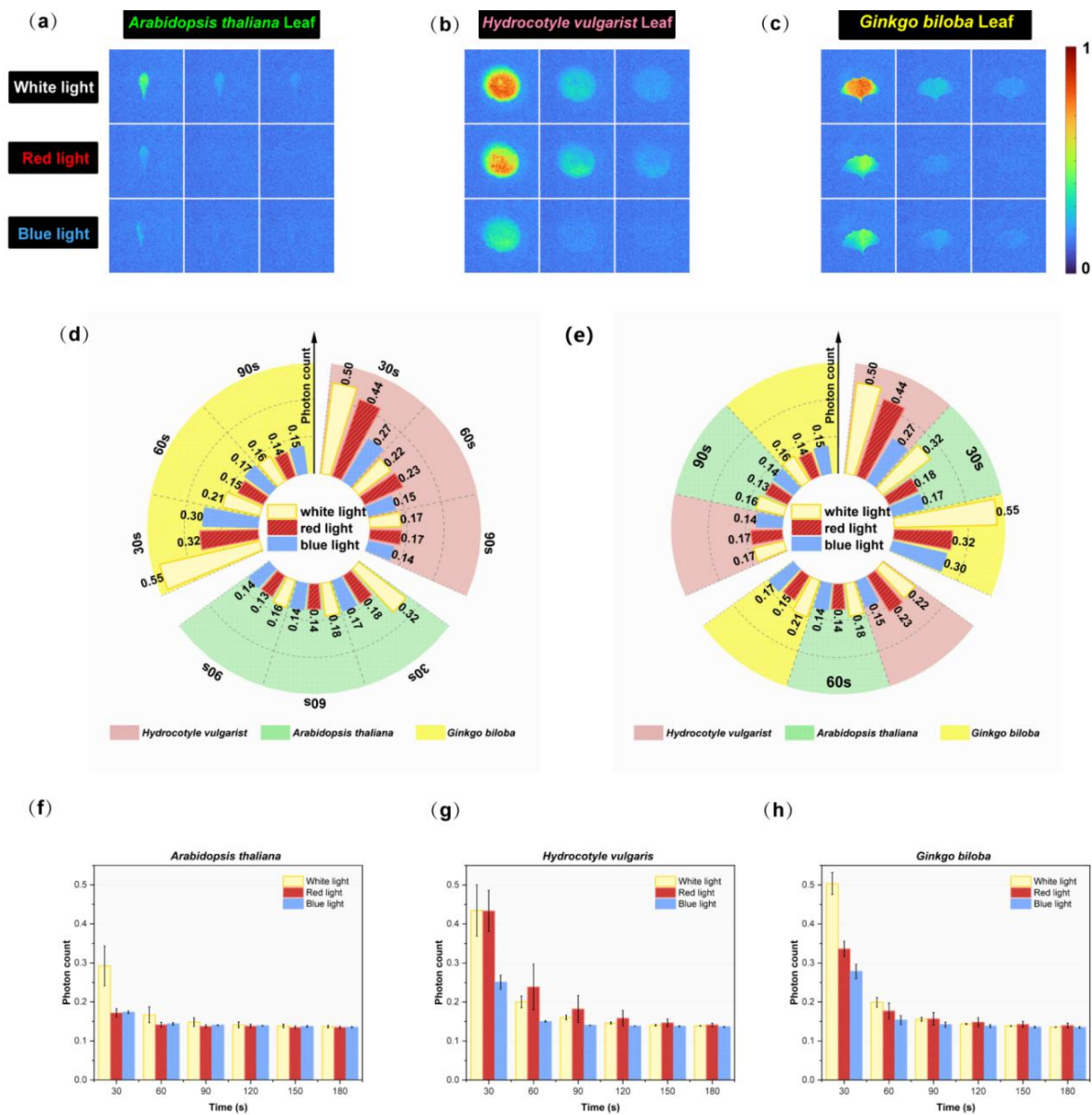

**Figure 3. Wavelength-dependent regulation of delayed luminescence (DL).** (a–c) Pseudo-color heatmaps showing spatial DL intensity distributions of the leaves of *Arabidopsis thaliana*, *Hydrocotyle vulgaris*, and *Ginkgo biloba* under white, red, and blue light (image resolution: 288 × 288 pixels, achieved by binning N=8 pixels). (d) DL intensity decay trends over time in each species at different wavelengths, corresponding to the photon counts derived from images (a-c). (e) Comparative DL intensity decay trends in the three plant species at identical exposure times at different wavelengths, corresponding to the photon counts derived from images (a-c). (f–h) Overall intensity decay trends for *Arabidopsis thaliana*, *Hydrocotyle vulgaris*, and *Ginkgo biloba* leaves at different wavelengths. Error bars represent standard deviation from three experimental replicates.

## 3.3 Stress-Specific Spatiotemporal Responses

3.3.1 Spatiotemporal Heterogeneity of DL Induced by Mechanical Stimulation

Using DL imaging to track DL in mechanically stimulated plant leaves, we investigated spatiotemporal DL responses following induction by mechanical stimulation. Next, we used the qCMOS system to capture DL intensity at sites of leaf injury under full resolution (2304 × 2304 pixels), as illustrated in **Figure 4a–b**. Imaging results for the initial 30 seconds are presented in **Figure 4a**, while those for the first 180 seconds are shown in **Supplementary Figure 1**.

Spatially distributed pseudo-color heatmaps (**Figure 4a**) revealed species-specific patterns of DL signal diffusion at injury sites post-mechanical stimulation. Specifically, we found that *Arabidopsis thaliana* leaves showed enhanced DL signals propagating directionally along the veins of the leaf, with significantly higher intensities observed in vein regions than in mesophyll tissue. In contrast, *Hydrocotyle vulgaris* leaves exhibited a radial DL enhancement pattern centered at the site of the wound, with diffusion pathways matching incision geometry. Finally, *Ginkgo biloba* leaves displayed isotropic DL diffusion around the injury site, covering a broader peripheral area.

Next, multiscale validation (**Figure 4b**) demonstrated the technical superiority of the qCMOS system for detecting injury-induced DL. At full resolution (N=1, 2304 × 2304 pixels), control *Hydrocotyle vulgaris* leaves exhibited uniformly low-intensity DL signals, while the 5-min mechanically damaged group showed a distinct radial enhancement pattern, with submillimeter DL hotspots at the edges of the wound. As the binning factor increased (i.e., N=2→4→8), the effective signal-to-noise ratio was maintained via equivalent pixel size expansion. Even in low-resolution mode (N=8, 288 × 288 pixels), the injury features remained prominent and showed enhanced sensitivity to weak DL photons.

Quantitative analysis of DL intensity 5 minutes post-injury revealed distinct interspecific response patterns (**Figure 4c–e**). *Hydrocotyle vulgaris* exhibited an 18.1% increase in initial DL intensity (0.539 $\pm$ 0.102 vs. control 0.457 $\pm$ 0.039 photons/pixel/30s), consistent with its radial diffusion pattern. *Ginkgo biloba* showed a more pronounced 27.3% enhancement (0.494 $\pm$ 0.107 vs. 0.388 $\pm$ 0.018 photons/pixel/30s), aligning with sustained isotropic diffusion. In contrast, *Arabidopsis thaliana* displayed no significant change (0.281 $\pm$ 0.004 vs. 0.242 $\pm$ 0.050 photons/pixel/30s), reflecting its rapid vein-localized response and efficient ROS scavenging.

Temporal dynamic analysis (**Figure 4f–h, derived from DL images in Figure 4a**) further confirmed species-specific DL intensity decay patterns post-mechanical stimulation. For example, *Arabidopsis thaliana* leaves demonstrated a "delayed peak response," with standardized intensity reaching 124.1% (i.e., 0.243 vs. 0.195 photons/pixel/30s) after 5 minutes. Peak intensity (0.321 photons/pixel/30s, representing a 64.4% increase over controls) occurred after 15 minutes, followed by rapid decay to 101.2% residual intensity (0.137 vs. 0.136 photons/pixel/30s) by 180 s. Moreover, *Hydrocotyle vulgaris* leaves displayed a "transient burst response," peaking at 5 minutes (0.653 photons/(pixel·30s) with a 188.7% increase over the control (0.346 photons/pixel/30s). This was followed by a rapid decline to 98.5% of the control level (i.e., 0.145 vs. 0.147 photons/pixel/30s) by 180 s. *Ginkgo biloba* leaves exhibited a "sustained enhancement response," with intensity rising to 162.8% after 5 minutes (0.596 vs. 0.366 photons/pixel/30s), a slower decay, and residual intensity remaining at 101.2% (0.139 vs. 0.137 photons/pixel/30s) by 180 s. After 30 minutes, DL intensity at the injury site remained at 136.4% of the control (0.496 vs. 0.366 photons/pixel/30s).

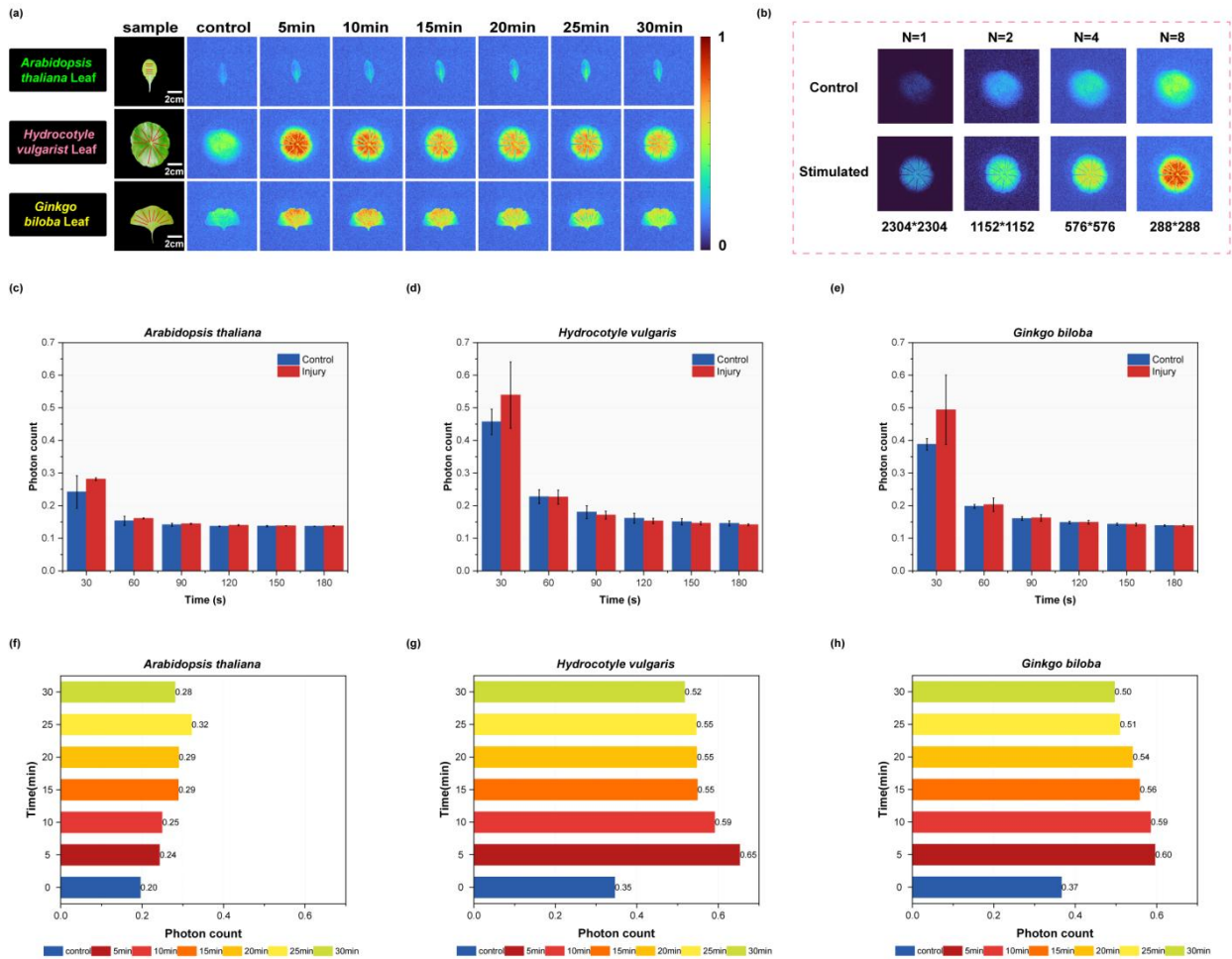

**Figure 4**. **Spatiotemporal responses of plant delayed luminescence (DL) following mechanical stimulation.** (a) Pseudo-color heatmaps showing initial DL intensity (30 s) distributions at injury sites (red-outlined regions) and adjacent areas at 5–30 minutes post-stimulation (representative images from a single exploratory experiment, scale bar = 2 mm; image resolution: 288 × 288 pixels). (b) Image processing results for *Hydrocotyle vulgaris* leaves. Control and mechanically damaged groups (5 min post-stimulation) under varying binning factors (N=1, 2, 4, 8), with labeled resolutions (e.g., "2304×2304"). (c–e) DL intensity comparisons between control and mechanically damaged groups (5 min post-stimulation) for *Arabidopsis thaliana* (c), *Hydrocotyle vulgaris* (d), and *Ginkgo biloba* (e) leaves. Histograms show photon counts with error bars representing standard deviation from three experimental replicates. （f–h)Temporal changes in initial DL photon counts (first 30 s) at injury sites: Control (0 min) and 5–30 minutes post-stimulation for *A. thaliana* (f), *H. vulgaris* (g), and *G. biloba* (h). Histograms display seven time points: control + six post-damage intervals (5, 10, 15, 20, 25, 30 min). Data are derived directly from the images shown in (a).

3.3.2 Species-Specific DL Responses Under Oxidative Stress

To further explore the impact of oxidative stress on DL, we analyzed the spatiotemporal DL responses in leaves treated with 3% $H_2O_2$ (**Figure 5**). Imaging results for the initial 30 seconds are presented in **Figure 5a**, and imaging results for the first 180 seconds are shown in **Supplementary Figure 2.**

Next, pseudo-color heatmaps (**Figure 5a**) revealed distinct species-specific DL distributions in response to oxidative stress. For example, *Arabidopsis thaliana* leaves exhibited weak responses localized to secondary veins, with limited diffusion throughout the leaf. *Hydrocotyle vulgaris* leaves showed gradient DL diffusion from the leaf center outward, with pronounced central enhancement. Finally, *Ginkgo biloba* leaves displayed an "edge-to-center" pattern in which DL signals extended from the leaf margin toward the midvein.

At 30 minutes post-$H_2O_2$ treatment, *Hydrocotyle vulgaris* showed a 17.2% increase in DL intensity ($0.451 \pm 0.058$ vs. Control $0.385 \pm 0.019$ photons/pixel/30s), supporting its central-to-peripheral diffusion (**Figure 5c**). *Ginkgo biloba* exhibited a 30.9% enhancement ($0.596 \pm 0.048$ vs. $0.455 \pm 0.039$ photons/pixel/30s), corroborating persistent "edge-to-center" DL propagation (**Figure 5d**). Conversely, *Arabidopsis thaliana* displayed significant suppression ($0.196 \pm 0.037$ vs. $0.238 \pm 0.059$ photons/pixel/30s), aligning with its vein-restricted response(**Figure 5b**).

Standardized DL intensity analysis (expressed relative to a negative control) also demonstrated species-specific decay trends (**Figure 5e–g, calculated from images in Figure 5a**). Here, *Arabidopsis thaliana* leaves showed transient suppression (i.e., 12.1% reduction at 30 s, standardized to 88%) in the 15-min treatment group, with recovery to the baseline by 90 s. Residual intensity at 30 minutes differed by <5% from the negative control, showing the fastest decay rate. In *Hydrocotyle vulgaris* leaves, the 5-minute $H_2O_2$ treatment group showed no significant increase in DL by 30 s, but intensities remained elevated post-60 s (i.e., +5.1% at 180 s). The 30-min group exhibited a biphasic response, featuring an initial reduction in DL (i.e., 91% at 30 s) followed by a late-phase increase in DL (i.e., 103% at 180 s). *Ginkgo biloba* leaves displayed sustained enhancement in response to oxidative stress, peaking at 162% standardized intensity at 30 s in the 25-min treatment group, with residual intensity maintained at 102% by 180 s, indicating that oxidative stress exerted persistent effects.

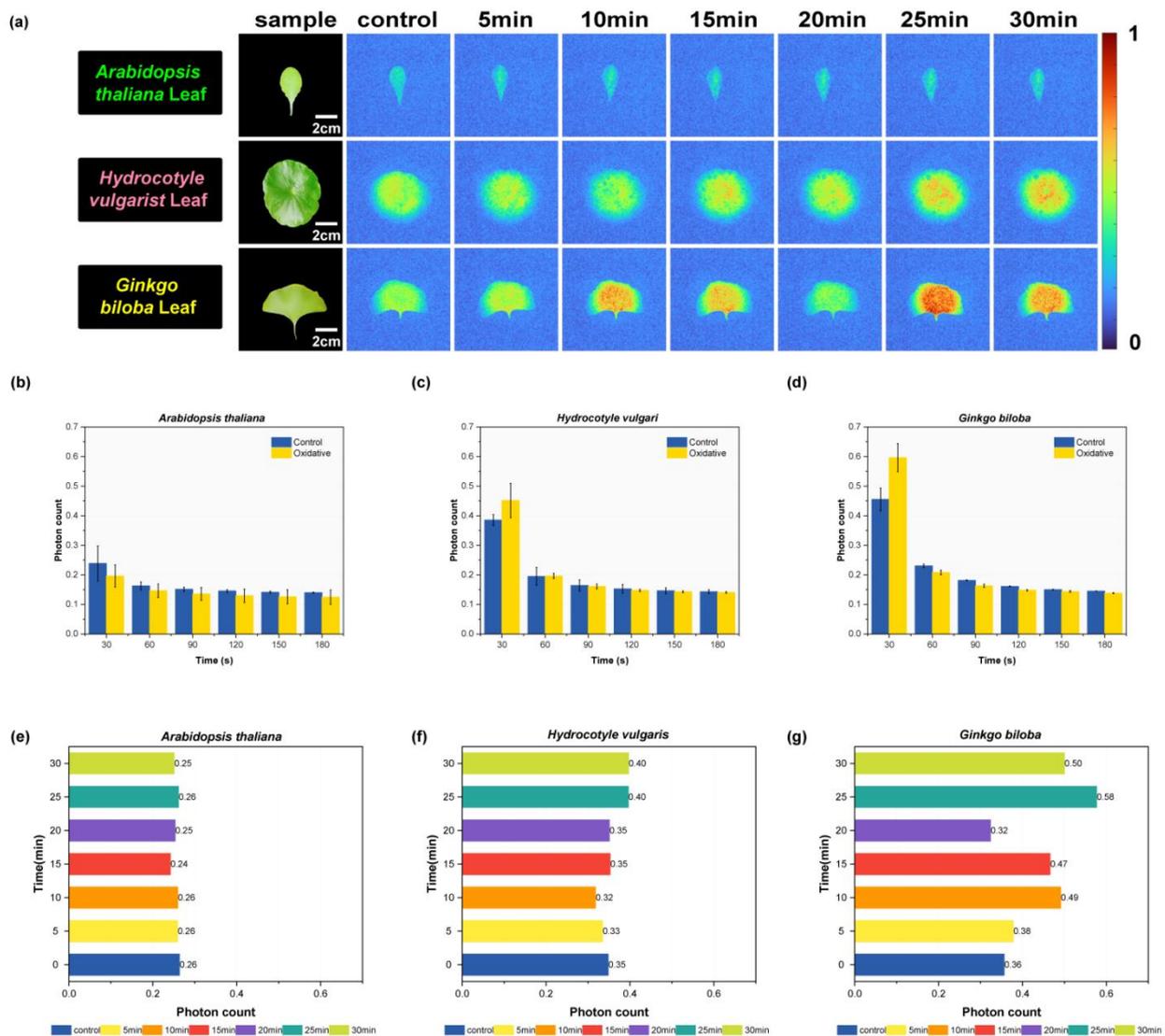

**Figure 5. Delayed luminescence (DL) responses to oxidative stress.** (a) Pseudo-color heatmaps showing spatial DL intensity distributions (30 s) in leaves of Arabidopsis thaliana, Hydrocotyle vulgaris, and Ginkgo biloba 5–30 minutes post-treatment with 3% $H_2O_2$ (representative images from a single exploratory experiment, scale bar = 2 mm; imaging size: 288 × 288 pixels). (b–d) DL photon counts comparing control and 30 min post-$H_2O_2$ treatment groups for A. thaliana (b), H. vulgaris (c), and G. biloba (d). Histograms show mean photon counts with error bars representing standard deviation from three experimental replicates. (e–g) Initial DL photon counts (first 30 s) in control (0 min) and 5–30 minutes post-$H_2O_2$ treatment (at 5 min intervals) for A. thaliana (e), H. vulgaris (f), and G. biloba (g). Time-course data derived from Figure 5a.

## 3.4 Two-Level Quantum Model

Traditional theoretical frameworks have typically employed discrete radical reaction chains to describe the processes related to plant ROS metabolism. In general, these theories excel in microscale chemical kinetic characterization[3,4,7]. However, significant challenges arise when integrating spatiotemporal heterogeneity in response to localized ROS bursts that can be induced by mechanical stimulation, gradient diffusion of oxidative stress, or wavelength-specific modulation of excited-state electron relaxation. In particular, conventional models often fail to establish quantitative correlations between DL decay dynamics and ROS metabolism due to inadequate characterization of the quantum properties of photoexcited electrons[1]. This limitation can create theoretical barriers that prevent cross-scale analyses designed to link plant photophysical processes to biochemical responses. To overcome these challenges, this study proposes a two-level quantum model, with the general aim of establishing a quantitative mapping between ROS metabolic chemical potential and DL kinetic parameters via a simplified quantum framework. We hope that this will facilitate an improved understanding of the physical essence of photoexcited state relaxation and thereby offer a theoretical tool for the systematic analysis of plant multifactorial stress responses.

Based on these objectives, our model simplifies the complex ROS metabolism into a chemical potential-driven two-level quantum system (**Figure 6**). Here, the ground state (g) represents unexcited electronic states, while the excited state (e) corresponds to photon-emitting excited states. The intermediate state (c) describes the dynamic equilibrium between ROS generation (i.e., chemical potential gain ($\Gamma_t^+$)) and antioxidant scavenging ($\Gamma_t^-$). The governing kinetic equations are:

$$\partial_t N_e = -\gamma N_e + \kappa N_c \qquad (1)$$

$$\partial_t N_c = -\Gamma_t^- N_c + \Gamma_t^+ N_g \qquad (2)$$

Here, the parameter $\Gamma_t^+$ integrates the thermodynamic driving force of ROS-generating reactions, scaling proportionally with local ROS concentration and metabolic flux. Crucially, $\Gamma_t^+$ intrinsically encodes spatiotemporal heterogeneity: spatial gradients (e.g., elevated $\Gamma_t^+$ in *Arabidopsis thaliana* veins or $H_2O_2$-diffused zones in *Hydrocotyle*) map directly onto DL intensity distributions. Temporally, $\Gamma_t^+$ spikes instantaneously at wound sites (e.g., 188.7% DL surge in *Hydrocotyle vulgaris*; **Figure 4g**) but rises progressively during $H_2O_2$ diffusion (e.g., *Hydrocotyle* vulgaris; **Figure 5f**). This phenomenological approach bypasses molecular complexity while retaining predictive power, as $\Gamma_t^+$ correlates DL intensity with ROS flux without requiring explicit Gibbs free energy calculations. Meanwhile, $\gamma$ (decay rate) reflects antioxidant scavenging efficiency (e.g., high $\gamma$ in *Arabidopsis thaliana*), and $\kappa$ (injection efficiency) captures tissue-dependent ROS diffusion (e.g., low $\kappa$ in thick *Ginkgo biloba* leaves).

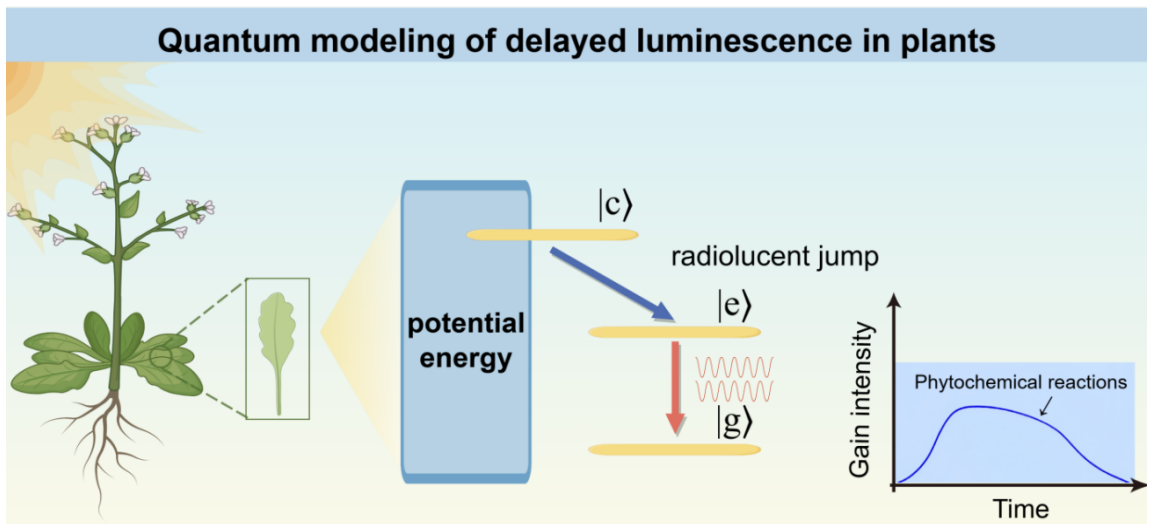

**Figure 6. Physical mechanism and a two-level quantum model of plant delayed luminescence (DL).** The energy levels e and g represent excited and ground states, respectively. Left panel: Chemical potential provided by complex ROS reactions. Curve: Temporal evolution of the chemical potential gain ($\Gamma_t^+$).

Relative to traditional theoretical frameworks, this model abstracts discrete ROS reactions (e.g., $^1O_2$ and HO· generation) into $\Gamma_t^+$-driven continuous processes, thereby bypassing molecular-level complexity. In doing so, it resolves DL spatiotemporal heterogeneity via $\Gamma_t^+$ spatial gradients, incorporates the wavelength-dependent value $\gamma$ to explain differential DL decay under white vs. monochromatic light, and quantifies the evolutionary specificity in DL kinetics by quantifying interspecies variation in $\gamma$ and $\kappa$. Overall, this framework retains the physical essence of radical reactions while enabling quantitative spatiotemporal resolution via parameterization.

To validate this model, we systematically analyzed DL responses under diverse stress conditions using qCMOS imaging data and kinetic equations. In *Hydrocotyle vulgaris* leaves, DL intensity surged to 188.7% of the value of the negative control (i.e., 0.65 vs. 0.35 photons/pixel/30s, **Figure 4c**) 5 minutes after mechanical stimulation. This corresponds to instantaneous $\Gamma_t^+$ activation and confirms this model's ability to capture ROS burst-driven electron injection. *Ginkgo biloba* leaves exhibited stepwise DL enhancement in response to oxidative stress (**Figure 5c**), reflecting multiphase $\Gamma_t^+$ activation (e.g., progressive ROS accumulation via $H_2O_2$ diffusion). With respect to light quality regulation, *Ginkgo biloba* displayed the highest initial DL intensity under white light (i.e., 0.50 photons/pixel/30s) and the fastest decay (26.9% residual vs. 41.6% under red and 48.5% under blue light), a finding that is consistent with $\gamma_{white} > \gamma_{monochromatic}$.

As a phenomenological theory, this model avoids explicitly accounting for specific radical species or enzymes and aims to unify experimental observations using parametric methods. Moreover, the spatial distribution of variation in DL (e.g., vein-specific signals in the leaves of *Arabidopsis thaliana*) directly map onto excited-state electron occupancy probabilities, pioneering a link between quantum effects and plant tissue architecture. This model therefore provides a unified framework for multifactorial stress responses, including mechanical stimulation (local $\Gamma_t^+$ spikes), oxidative stress (global $\Gamma_t^+$ activation), and light quality modulation (selective $\gamma$ tuning). In general, $\Gamma_t^+$ serves as a stress sensitivity index, revealing universal ROS metabolic principles. Moreover, interspecies variation in $\gamma$ and $\kappa$ elucidate evolutionary adaptations in thick-walled vs. thin-leaved plants. By modulating monochromatic light to reduce $\gamma$, the DL lifetime can be extended to enhance crop stress resistance.

## 4. Discussion

This study established a high-resolution imaging system based on qCMOS technology. In doing so it enables the spatiotemporal dynamic analysis of DL signals in plants in response to mechanical stimulation, oxidative stress, and differences in light quality. This methodological advance provides novel perspectives and a powerful technical tool for plant stress physiology research[25-32]. Leveraging this innovative technology, we demonstrated the spatiotemporal specificity and interspecies heterogeneity of plant DL responses to stress, further characterizing a relationship between such response mechanisms and specific plant stress-resistance strategies.

The qCMOS-based high-resolution DL imaging system developed here overcomes the limitations of traditional plant stress detection technologies regarding spatiotemporal resolution and noninvasive monitoring, and offers a methodological advancement for plant photobiology. Conventional stress studies often rely on destructive biochemical assays (e.g., ROS fluorescent probes or enzyme activity measurements), which yield static molecular-level data but cannot capture real-time spatiotemporal dynamics induced by physiological perturbations. In contrast, our system achieves noninvasive DL signal acquisition with a temporal resolution of 30 seconds and a maximum spatial resolution of 2304 × 2304 pixels, thereby facilitating continuous tracking of DL spatial diffusion trajectories for up to 30 minutes post-stress. Moreover, this system has a single-photon resolution (readout noise ≤0.3 e⁻/pixel) and a broad spectral range (400–1000 nm), which supports wavelength-specific DL detection under red (660 nm) and blue (460 nm) light. In addition, pixel binning (N=1→8) enhances the clarity of stress or injury signal diffusion patterns, balancing the resolution and the signal-to-noise ratio. The results corresponding to different binning factors are shown in Supplementary Figures 3-5.

We used this system to observe interspecies differences in DL responses to mechanical stimulation. We found that *Hydrocotyle vulgaris* leaves exhibited a "transient burst" DL pattern characterized by a sharp increase in DL intensity at 5 minutes post-stimulation; this burst is likely linked to rapid ROS diffusion within its thin-walled leaf structure[33]. In contrast, *Ginkgo biloba* leaves displayed a "sustained enhancement" response, potentially due to physical constraints in ROS diffusion imposed by thick-walled tissues and antioxidative buffering[34]. *Arabidopsis thaliana* leaves showed rapid recovery after reaching peak DL intensity 15 minutes post-stimulation, suggesting efficient ROS scavenging[35]. These differential responses indicate that DL kinetics are closely associated with interspecific variation in leaf anatomy (e.g., mesophyll thickness) and antioxidant systems[36,37]. By quantifying DL decay rates (e.g., slower decay phases in *Ginkgo biloba* leaves relative to *Arabidopsis thaliana* leaves), this study identifies potential indicators for stress-resistance phenotyping.

Next, oxidative stress experiments (**Figure 5**) also revealed species-specific patterns in DL spatial distribution. For example, *Hydrocotyle vulgaris* leaves displayed gradient diffusion from the center outward, *Ginkgo biloba* leaves showed sustained enhancement, and *Arabidopsis thaliana* leaves showed responses localized to the secondary veins. These variations likely reflect tissue-specific patterns of antioxidative system organization[38]. The "edge-to-center conduction" pattern in *Ginkgo biloba* leaves may be related to the epidermal barrier properties such that cuticle-mediated $H_2O_2$ release establishes a persistent oxidative microenvironment, while flavonoid-associated energy storage is released via DL[39]. Importantly, this imaging method provides a noninvasive approach for dynamically monitoring oxidative stress and reveals species-specific antioxidative strategies through distinct luminescence signatures. Relative to the mechanical stimulation experiments, our system demonstrated higher temporal sensitivity during the oxidative stress trails (e.g., 61.9% intensity change in *Ginkgo biloba* leaves by 25 minutes), as well as notable differences in spatial patterns (e.g., central diffusion in *Hydrocotyle vulgaris* vs. radial distribution following injury). Among them，*Ginkgo biloba* leaves exhibited maximal DL intensity at 25 min post-$H_2O_2$ treatment (162% standardized intensity), which may be attributed to: (1) Progressive ROS accumulation via cuticle-limited $H_2O_2$ diffusion, establishing persistent oxidative microenvironments[39]; (2) Thick mesophyll physically retaining ROS and prolonging excited-state electron lifetimes[34]; (3) Flavonoid-mediated antioxidative buffering delaying radical termination[34,39]. Significantly, this temporal trend was validated by triplicate biological replicates at the 30-min endpoint (**Figure 5**d), showing a 30.9% DL enhancement (0.596 ± 0.048 vs. Control 0.455 ± 0.039 photons/pixel/30s). Future studies could integrate ROS fluorescent probes to validate the biological applicability of this technique.

Subsequent light quality experiments further elucidated the previous finding of wavelength-dependent DL modulation. We found that white light induced the highest initial DL intensity (e.g., 0.50 photons/pixel/30s in *Ginkgo biloba* leaves) but also the fastest decay (26.9% residual at 180 s). Moreover, red and blue light differentially stabilized DL; red light induced slower decay (e.g., 41.6% residual in *Ginkgo biloba* leaves), whereas blue light maintained high stability (e.g., >77% residual in *Arabidopsis thaliana* leaves). Red light likely stabilizes ROS-scavenging pathways, prolonging DL persistence, while blue light accelerates ROS generation, enhancing initial DL but hastening decay. Collectively, these wavelength-dependent regulatory effects likely arise from light signaling pathways that modulate ROS metabolism and/or electron transport chain activity[35,40]. Wavelength-dependent DL responses, such as the prolonged DL lifetime observed in *Hydrocotyle vulgaris* under red light, offer new insight into spectral management for controlled-environment agriculture.

The two-level quantum model developed in this study provides a unified theoretical framework for analyzing DL spatiotemporal dynamics[41]. ROS-driven mechanisms underlie species-specific DL variation. Species adaptations manifest in model parameters: $\Gamma_t^+$ (ROS chemical potential) scales with stress intensity, explaining DL surges in wounded *Hydrocotyle vulgaris* (188.7% increase, **Figure 4c**). $H_2O_2$-induced DL amplification across species (**Figure 5**) confirms $\Gamma_t^+$ as the central regulator of stress-responsive DL. *Arabidopsis thaliana*' high $\gamma$ (rapid decay) reflects efficient enzymatic ROS scavenging, while *Ginkgo biloba*'s low $\kappa$ (injection efficiency) arises from physical ROS retention in thick mesophyll. This phenomenological mapping avoids unmeasurable thermodynamic potentials yet directly links quantum relaxation (excited-state electron occupancy) to macroscopic ROS gradients. Thin-leaved species (*Hydrocotyle vulgaris*) permit rapid ROS dispersion, causing transient DL bursts at injury sites (**Figure 4d**). In contrast, thick-leaved *Ginkgo biloba* retains ROS locally, sustaining DL emission (**Figure 4e**, **Figure 5g**). Through parameterization ($\gamma$, $\kappa$, $\Gamma_t^+$), this model mechanistically explains variations in DL response under mechanical stimulation, oxidative stress, and light quality regulation, and bypasses the added complexity of accounting for all molecular details. This model supplements existing theories regarding radical metabolism by directly mapping DL intensity distributions to excited-state electron occupation probabilities; it therefore establishes the first quantitative link between quantum effects and plant tissue architecture, and demonstrates potential for cross-scale plant stress research in the future.

The interspecies heterogeneity in DL kinetics is primarily attributed to divergent ROS management strategies and leaf anatomical adaptations. *Arabidopsis thaliana* veins exhibit localized ROS accumulation due to high metabolic activity, driving elevated DL intensity but rapid decay via efficient enzymatic scavenging (e.g., ascorbate peroxidases) [3,35]. Conversely, *Ginkgo biloba*'s thick mesophyll physically restricts ROS diffusion and leverages flavonoid-mediated antioxidative buffering, prolonging electron relaxation and DL persistence [34,39]. *Hydrocotyle vulgaris*, with minimal structural barriers, achieves rapid ROS equilibration, resulting in homogeneous DL distribution and intermediate decay rates [36]. These patterns confirm that spatial ROS gradients directly modulate DL spatiotemporal dynamics, as ROS-generated excited species (e.g., triplet carbonyls, singlet oxygen) serve as primary photon emitters [3,9].

Notably, ROS dynamics govern DL spatial patterns. Our data demonstrate three key lines of evidence dissociating DL from chlorophyll distribution: (1) Vascular bundles in Arabidopsis thaliana exhibit lower chlorophyll density yet elevated DL due to ROS-rich metabolic activity; (2) Mechanically wounded sites in Hydrocotyle vulgaris show peak DL despite chlorophyll structural disruption; (3) DL emission spectra (e.g., 707.8/730.3 nm) are red-shifted >25 nm relative to chlorophyll-a fluorescence (682 nm), confirming its origin in photosynthetic electron transport (PSI/II) energy loss pathways[15-17]. These patterns establish that DL spatial heterogeneity reflects ROS-driven photophysical processes, not chlorophyll distribution or pigment density—supporting our conclusion that species-specific DL kinetics are governed by stress-responsive redox states and electron transport efficiency.

Despite these advancements, unresolved questions remain. The experiments reported here focus on acute oxidative stress (e.g., mechanical stimulation and $H_2O_2$ treatment), but chronic stress-induced evolution in DL response requires further investigation. While our 180-second window captured the dominant DL dynamics, future studies should extend measurements for species with slower decay (e.g., woody plants[42]) to fully resolve asymptotic behaviors. While baseline chlorophyll differences contribute to interspecies DL intensity variations, our acute stress results (<30 min) reflect ROS-driven photophysics rather than pigment concentration artifacts. Future work should correlate chlorophyll dynamics with DL during prolonged stress. In addition, the temperature dependence of specific model parameters (e.g., $\gamma$) remains uncharacterized, necessitating temperature-controlled experiments to further refine this framework. Moreover, the limited detection efficiency in the near-infrared range (>900 nm) hinders the analysis of quantum effects in deep tissues. Future studies should enhance device performance and expand experimental scope to improve the spatiotemporal monitoring of plant stress responses. In doing so they can further explore the potential of DL imaging for understanding photophysical regulation and improving stress-resistant crop breeding.

# 5. Conclusion

This study establishes a high-resolution qCMOS imaging platform (2304 × 4096 pixels, 30 s per frame) for spatiotemporal mapping of plant DL. Our system resolves spatial heterogeneity in DL intensity across species—revealing vein-localized signals in *Arabidopsis thaliana,* radial diffusion patterns in mechanically wounded *Hydrocotyle vulgaris*, and sustained "edge-to-center" propagation in $H_2O_2$-stressed *Ginkgo biloba*. Crucially, we demonstrate that DL kinetics are governed by species-specific adaptations: thin-leaved *Hydrocotyle vulgaris* exhibits rapid ROS diffusion and transient DL bursts, while thick-leaved *Ginkgo biloba* retains ROS locally, prolonging DL emission. *Arabidopsis thaliana* achieves rapid signal decay via efficient antioxidant scavenging. Furthermore, spectral experiments reveal wavelength-dependent modulation—white light maximizes initial DL intensity but accelerates decay, whereas red light stabilizes emission persistence (>77% residual in *Arabidopsis thaliana*)—providing actionable insights for spectral management in controlled agriculture.

To integrate these findings, we propose a two-level quantum model that parametrically links DL dynamics to fundamental ROS metabolic principles. The model quantifies stress responses through three key parameters: $\Gamma_t^+$ (ROS chemical potential) scales with stress intensity, $\gamma$ (decay rate) reflects antioxidant efficacy, and $\kappa$ (electron injection efficiency) captures tissue-dependent ROS diffusion. This framework bridges quantum relaxation events (excited-state electron occupancy) with macroscopic stress phenotypes, bypassing molecular complexity while preserving predictive power.

Collectively, our integrated approach advances plant photobiology by: (1) enabling noninvasive, real-time tracking of stress responses at submillimeter resolution; (2) establishing DL as a quantitative biomarker of interspecific stress resilience; and (3) providing a theoretical framework for optimizing crop stress resistance through spectral management and targeted antioxidant enhancement.

# 6. Abbreviations

| | |
|---|---|
| DL | delayed luminescence |
| qCMOS | quantitative complementary metal-oxide-semiconductor |
| UPE | ultra-weak photon emission |
| ROS | reactive oxygen species |
| PMT | photomultiplier |
| EMCCD | electron-multiplying charge-coupled device |
| ICCD | intensified charge-coupled device |
| sCMOS | scientific complementary metal-oxide-semiconductor |
| PNR | photon number resolution |
| SNR | signal-to-noise ratio |

# Acknowledgments

The authors would like to thank Figdraw (www.figdraw.com) for help with character cartoons.

# Author Contributions

**Yan-Xia Liu**: Conceptualization, Methodology, Software, Investigation, Formal Analysis, Validation, Writing- original draft. **Hai-Yu Fan**: Methodology, Investigation, Formal Analysis. **Yu-Hao Wang**: Software, Data curation. **Yan-Liang Wang**: Conceptualization, Methodology. **Sheng-Wen Li**: Conceptualization, Methodology, Investigation. **Shi-Jian Li**: Conceptualization, Methodology, Writing−review & editing. **Xu-Ri Yao**: Conceptualization, Formal analysis, Methodology, Resources, Writing−review & editing. **Qing Zhao**: Resources, Supervision, Writing−review & editing.

# Funding

This research was not specifically supported by specific grants from funding agencies from the public, commercial, or not-for-profit sectors.